\documentclass[submission,copyright,creativecommons]{eptcs}
\providecommand{\event}{LFMTP 2020} 
\usepackage{breakurl}             
\usepackage{underscore}           

\usepackage{amssymb}
\usepackage{amsmath}%
\usepackage{amsfonts}%
\usepackage{cmll} 
\usepackage{proof}
\usepackage{color}
\usepackage{enumitem}

\definecolor{proccolor}{rgb}{0.1,0.1,1.0}

\DeclareMathOperator{\tstile}{\ \vdash}

\DeclareMathOperator{\cln}{:}

\DeclareMathOperator{\munion}{\uplus}
\DeclareMathOperator{\mimp}{\Longrightarrow}
\DeclareMathOperator{\mempty}{\varnothing}

\newcommand{\msing}[1]{\{#1\}}
\newcommand{\hl}[1]{\texttt{#1}}
\newcommand{\chan}[1]{#1}

\newcommand{\seq}[2]{\textcolor{proccolor}{\chan{#1}\cln }#2}

\newcommand{\sep}{,\ }

\title{Object-Level Reasoning with Logics Encoded in HOL Light}
\author{Petros Papapanagiotou
\institute{School of Informatics\\
University of Edinburgh\\
Edinburgh, United Kingdom}
\email{ppapapan@inf.ed.ac.uk}
\and
Jacques Fleuriot
\institute{School of Informatics\\
University of Edinburgh\\
Edinburgh, United Kingdom}
\email{jdf@inf.ed.ac.uk}
}
\def\titlerunning{Object-Level Reasoning with Logics Encoded in HOL Light}
\def\authorrunning{P. Papapanagiotou \& J. Fleuriot}
\begin{document}
\maketitle

\begin{abstract}
We present a generic framework that facilitates object level reasoning with logics that are encoded within the Higher Order Logic theorem proving environment of HOL Light. This involves proving statements in any logic using intuitive forward and backward chaining in a sequent calculus style. It is made possible by automated machinery that take care of the necessary structural reasoning and term matching automatically. Our framework can also handle type theoretic correspondences of proofs, effectively allowing the type checking and construction of computational processes via proof. We demonstrate our implementation using a simple propositional logic and its Curry-Howard correspondence to the $\lambda$-calculus, and argue its use with linear logic and its various correspondences to session types.
\end{abstract}

\section{Introduction}%
\label{sec:introduction}

Higher order logic (HOL) proof assistants, such as HOL Light~\cite{harrison1996hol}, provide the means to encode other languages and prove conjectures specified within them (\emph{object-level reasoning}). This can help understand how the objects of an encoded logic behave or develop practical applications using that logic, such as the development of correct-by-construction programs, type checking, or verification of specific terms or programs (such as the work on linear logic in Coq by Power et al.\cite{power1999working}).

In practice, performing proofs within an encoding of a custom logic in HOL typically requires the development of specialised tools and tactics, in what can be seen as \emph{a theorem prover within a theorem prover}. These include, for example, mechanisms for fine-grained structural manipulation of the encoded formulae, customised application of the inference rules, seamless backward and forward chaining, etc. Users often need to develop such tools in an ad-hoc way in order to reason specifically about the logic they are interested in, and this drastically increases the effort and time required to obtain a useful encoding. Moreover, such tools may not scale across other logics, leading to a replication of effort. 

We present a generic framework and toolset for object-level reasoning with custom logics encoded in HOL Light. It aims to facilitate the exploration of different logic theories by minimizing the effort required between encoding the logic and obtaining a proof. Assuming a sequent calculus style encoding of the logic's inference rules, it allows their direct and intuitive application both backwards and forwards, while the system automates most of the structural reasoning required. The implemented tactics can also be used programmatically to construct automated proof procedures.



An important part of our framework, which sets it apart from similar systems like Isabelle (see Section~\ref{sec:related-work}), is the handling of type theoretical correspondences between encoded logics and programs. These are inspired by the \emph{propositions-as-types} paradigm (or \emph{Curry-Howard correspondence}) between intuitionistic logic proofs and the $\lambda$-calculus~\cite{Howard80}. They enable the construction of executable terms in some calculus via proof, also referred to as \emph{computational construction}. Terms are attached on each logical formula, so that the formula represents the type of the term. The application of inference rules on such a term annotated with its type, results in the construction of a more complex computational expression with some guaranteed correctness properties. Typically, cut-elimination in the proof corresponds to a reduction or execution step in the constructed term.

Our framework allows the user to easily construct computational terms and test and compare different logics and the behaviour of their correspondences (or even multiple correspondences of the same logic at the same time), merely by manipulating the encoding of the inference rules rather than having to rebuild or repurpose the implemented tools each time. More specifically, our framework allows 2 types of proofs:
\begin{itemize}
\item \emph{Type checking} proofs involve conjectures whose computational translation is given and the proof can only succeed if the conjecture is provable and the translation is correct. Such proofs effectively verify the types of programs.
\item \emph{Construction} proofs involve conjectures whose computational translation is unknown and is constructed as the proof progresses. Such proofs allow us to construct programs that match a given type via proof.
\end{itemize}

One of the main motivations of our work lies in the variety of computational translations between linear logic and session types, which we discuss briefly in Section~\ref{sec:further}.

\section{Example: Simple Logic}%
\label{sec:example:simple}

In order to provide simple examples of the functionality of our system, we focus on a subset of propositional logic involving only conjunction ($\times$) and implication ($\rightarrow$). 
More specifically, the terms of our example propositional logic can be defined as follows, using True ($\mathbf{T}$) as a constant/bottom object:
\begin{equation}
    \label{eq:syntax}
  Prop\ ::=\ \mathbf{T}\ |\ Prop\times Prop\ |\ Prop\rightarrow Prop 
\end{equation}

Our system is tailored to sequent calculus formulations, mainly because sequent calculus deduction is easier to encode as an inductive definition within HOL (and particularly in HOL Light). Note, however, that we use an intuitionistic sequent calculus which is equivalent to natural deduction~\cite{Girard:1989:PT:64805}. 

The inference rules of our simple logic are shown in Figure~\ref{fig:simplelogic}. Note that $\Gamma$ and $\Delta$ represent \emph{contexts}, i.e.\ finite sequences of formulae that are not affected by the application of the rule.

\begin{figure}[htbp]
  \centering
  \begin{tabular}{c}
    Identity \& Cut \\\hline
    \begin{minipage}{0.8\linewidth}
      \vspace*{3mm}
      \begin{equation}
        \nonumber
        \infer[Ax]{A \tstile A}{}
        \qquad\qquad
        \infer[Cut]{\Gamma \sep \Delta \tstile C}{
        \Gamma \tstile A
        &
        \Delta \sep A \tstile C
      }
      \end{equation}      
    \end{minipage}\vspace*{5mm}\\
    Structural Rules \\\hline
    \begin{minipage}{0.8\linewidth}
      \begin{equation}
        \nonumber
        \infer[Exchange]{\Gamma \sep B \sep A \sep \Delta \tstile C}{
        \Gamma \sep A \sep B \sep \Delta \tstile C
      }
        \qquad\qquad
        \infer[W]{\Gamma \sep A \tstile C}{
        \Gamma \tstile C
      }
        \qquad\qquad
        \infer[C]{\Gamma \sep A \tstile C}{
        \Gamma \sep A \sep A \tstile C
      }
      \end{equation}      
    \end{minipage}\vspace*{5mm}\\
    Logical Rules \\\hline
    \begin{minipage}{0.8\linewidth}
      \begin{equation}
        \nonumber
        \infer[L1\times]{\Gamma \sep A \times B \tstile C}{
        \Gamma \sep A \tstile C
      }
        \qquad\qquad
        \infer[L2\times]{\Gamma \sep A \times B \tstile C}{
        \Gamma \sep B \tstile C
      }
        \qquad\qquad
        \infer[R\times]{\Gamma \sep \Delta \tstile A \times B}{
        \Gamma \tstile A
        &
        \Delta \tstile B
      }
    \end{equation}\vspace*{-5mm}
    \begin{equation}
        \nonumber
        \infer[L\rightarrow]{\Gamma \sep \Delta \sep A\rightarrow B \tstile C}{
        \Gamma \sep B \tstile C
        &
        \Delta \tstile A
      }
        \qquad\qquad
        \infer[R\rightarrow]{\Gamma \tstile A\rightarrow B}{
        \Gamma \sep A \tstile B
      }
      \end{equation}      
    \end{minipage}

		\end{tabular}
			\caption{The inference rules of the subset of propositional logic used as an example.}
			\label{fig:simplelogic}
\end{figure}

In the next sections we investigate how this example logic can be encoded in HOL Light and then how our framework enables object-level proofs automatically without any further coding.

\subsection{Encoding}%
\label{sec:embedding}

A standard way of encoding such a logic in HOL Light starts by defining the syntax of the terms, in our case as shown in \eqref{eq:syntax}, as an inductive type. Note that, for clarity, we typeset definitions and formulas using the standard logical symbols for the various connectives instead of the harder to read ASCII-based syntax of HOL Light. The actual HOL Light implementation is available in the code base (see Section~\ref{sec:implementation}) and we also point out mappings between standard logic and HOL Light syntax where necessary.

We then need to provide the means to describe a sequent based on logical consequence (denoted by $\tstile$). A sequent may contain multiple formulas, including multiple copies of the same formula (such as in the contraction rule $C$ in Figure~\ref{fig:simplelogic}), and is thus often represented using a list of terms. 

Like most logics, our example includes an $Exchange$ rule (see Figure~\ref{fig:simplelogic}). This means that the order of the terms in the sequent is not important, so that $A \sep B\tstile C$ and $B\sep A \tstile C$ are semantically identical. For this reason, in paper proofs the $Exchange$ rule is almost always used implicitly, especially in complex proofs which may require a large number of applications of $Exchange$ to manipulate the sequents in the right form. We avoid their use altogether in our framework by replacing lists of formulas in a sequent with multisets\footnote{Sets may also be used to eliminate the need for weakening and contraction rules, but we prefer multisets as a more general approach that allows us to encode Linear Logic (see Section~\ref{sec:further}).}. In this case, the mechanism of bringing the sequents in the appropriate form relies on reasoning about and matching multisets, and these tasks can be automated more efficiently (see Section~\ref{sec:multisets})\footnote{However rare, non-commutative (or non-associative) substructural logics do exist, such as Lambeck's calculus of syntactic types~\cite{lambek1961calculus}. These can still be expressed in our framework using lists and will work with our tactics, but will simply not take advantage of the specialised multiset matching.}. 

An existing theory in HOL Light already has many useful formalised properties of multisets, and we added some more in our own small library extension. A variety of methods and tools, such as multiset normalisation, needed for our particular tasks were also implemented. Finally, we introduced HOL Light abbreviations for multiset sums (\hl{\^}) and singleton multisets (\hl{'}), in order to obtain a cleaner syntax. In this paper though, for readability, we use the $\munion$ symbol to denote a multiset sum, whereas enumerated multisets are enclosed in curly brackets $\msing{\cdot}$.

Based on the above, the sequent $\Gamma\sep A\tstile B$ can be represented using $\tstile$ as an (infix) boolean function in the term $(\Gamma \munion \{A\})\tstile B$. Note that a context $\Gamma$ is represented by a \emph{multiset variable}, i.e.\ a variable of type $(Prop)multiset$, whereas formulas $A$ and $B$ are term variables of type $Prop$. Also note that other logics may have additional or different types of arguments. For example, a regular (non-intuitionistic) sequent calculus will need 2 multisets of terms (left and right side) and some formulations of linear logic may need 4 multisets of terms (to distinguish linear from non-linear contexts). Our framework can deal with any of these situations with no additional specification other than the definition of $\tstile$ as a function.

Using this representation of sequents, the inference rules of the logic can be encoded as an inductive definition of logical consequence $\tstile$ as a function \hl{|--} in HOL Light. Note that this is different from HOL Light's syntax for entailment in proven HOL theorems \hl{|-}, which we omit here altogether to prevent any ambiguity. In the encoding, HOL implication ($\mimp$) expresses valid derivations between sequents, making HOL the effective meta-logic. As an example, the $R\times$ rule from Figure~\ref{fig:simplelogic} can be specified in the definition of $\tstile$ as follows:
\begin{equation}
  \label{eq:XR}
  \Gamma \tstile A \mimp \Delta \tstile B \mimp (\Gamma \munion \Delta) \tstile A\times B  
\end{equation}


\subsection{Object-level proofs}%
\label{sec:proofs}

So far we covered a straightforward encoding of a logic in HOL Light, with tools that are already available. Given such an encoding, one would expect to be able to do some object-level proofs. Let us take the commutativity of conjunction ($\times$), i.e.\ $\tstile X\times Y \rightarrow Y\times X$ as an example of such a proof:
\begin{equation}
  \label{eq:timescomm}
  \infer[R\rightarrow]{\tstile X\times Y \rightarrow Y\times X}{
    \infer[C]{X\times Y \tstile Y\times X}{
      \infer[R\times]{X\times Y\sep X\times Y \tstile Y\times X}{
        \infer[L1\times]{X\times Y\tstile X}{
          \infer[Ax]{X\tstile X}{}
        }
        &
        \infer[L2\times]{X\times Y\tstile Y}{
          \infer[Ax]{Y\tstile Y}{}
        }
      }
    }
  }
\end{equation}

Using vanilla HOL Light, proof~\eqref{eq:timescomm} can be accomplished using the interactive steps shown in Figure~\ref{fig:hlproof}. The actual proof script is available online and executed interactively (see link in Section~\ref{sec:implementation}).

\begin{figure}[htbp]
  \centering
  \begin{tabular}{|r|l|c|}\hline
    Step & Tactic & Goal(s) \\\hline
         & \emph{Initial state} & $\mempty\tstile X\times Y \rightarrow Y\times X $ \\
    1 & \hl{MATCH_MP_TAC ($R\rightarrow$)} & $\mempty\munion \msing{X\times Y} \tstile Y\times X $ \\
    2 & \hl{MATCH_MP_TAC ($C$)} & $\mempty\munion (\msing{X\times Y}\munion \msing{X\times Y}) \tstile Y\times X $ \\
    3 & \hl{REWRITE_TAC[MUNION_EMPTY2]} & $\msing{X\times Y}\munion \msing{X\times Y} \tstile Y\times X $ \\
    4 & \hl{MATCH_MP_TAC ($R\times$)} & $\msing{X\times Y} \tstile Y \ \wedge\ \msing{X\times Y} \tstile X $ \\
    5 & \hl{CONJ_TAC} & $\msing{X\times Y} \tstile Y \qquad \msing{X\times Y} \tstile X $ \\
    6 & \hl{SUBGOAL_THEN ($\mempty \munion \msing{X \times Y} \tstile Y$)...} & $\mempty\munion\msing{X\times Y} \tstile Y \qquad \msing{X\times Y} \tstile X $ \\
    7 & \hl{MATCH_MP_TAC ($L2\times$)} & $\mempty\munion\msing{Y} \tstile Y \qquad \msing{X\times Y} \tstile X $ \\
    8 & \hl{REWRITE_TAC[MUNION_EMPTY2]} & $\msing{Y} \tstile Y \qquad \msing{X\times Y} \tstile X $ \\
    9 & \hl{MATCH_ACCEPT_TAC ($ID$)} & $\qquad\qquad \msing{X\times Y} \tstile X $ \\
    10 & \hl{SUBGOAL_THEN ($\mempty \munion \msing{X \times Y} \tstile X$)...} & $\qquad \qquad \mempty\munion\msing{X\times Y} \tstile X $ \\
    11 & \hl{MATCH_MP_TAC ($L1\times$)} & $\qquad \qquad \mempty\munion\msing{X} \tstile X$ \\
    12 & \hl{REWRITE_TAC[MUNION_EMPTY2]} & $\qquad \qquad \msing{X} \tstile X$ \\
    13 & \hl{MATCH_ACCEPT_TAC ($ID$)} &  \\\hline
  \end{tabular}
  \caption{Interactive steps in HOL Light performing proof~\eqref{eq:timescomm}.}
  \label{fig:hlproof}
\end{figure}

Although there are 7 rule applications in the original proof, the HOL Light script contains almost double the steps even for such a simple, straightforward proof. 

The main tactic that allows the application of inference rules is \hl{MATCH\_MP\_TAC}, which matches the consequent of a HOL implication in a theorem with the goal and sets the antecedent as the new goal. Our rules are expressed using implication, so \hl{MATCH_MP_TAC} fits well for this use. However, the consequent and goal must match exactly for the tactic to work. This forces us to constantly manipulate the structure of the goal to ensure the rule is applied correctly. 

Step $1$ is straightforward. Observing the $R\rightarrow$ rule in Figure~\ref{fig:simplelogic} and the initial state, it is clear that $\Gamma$ in the rule is matched to the left-hand side of the turnstile, i.e.\ $\mempty$. The new goal is produced by the antecedent of the rule $\Gamma\sep A\tstile B$, where the left-hand side becomes $\Gamma\munion\msing{A}$, i.e.\ $\mempty\munion\msing{X\times Y}$. The empty multiset is therefore carried forward in the new goal.

This causes a problem after Step $2$, where we want to apply the $R\times$ rule. If we were to immediately apply \hl{MATCH_MP_TAC}, it would yield the instantiation $\Gamma = \mempty$ and $\Delta = \msing{X\times Y}\munion \msing{X\times Y}$. However, we would rather have a different context split, i.e.\ one where  $\Gamma =  \msing{X\times Y}$ and $\Delta = \msing{X\times Y}$. In this particular case, this is easily solved by eliminating the empty multiset via rewriting with the theorem \hl{MUNION_MEMPTY2}\footnote{\url{https://github.com/PetrosPapapa/hol-light-tools/blob/91f3f1b030728e0edf3ec86732d185207839a8ad/Library/multisets.ml\#L75}}, i.e.\ the property $\forall M.\mempty\munion M = M$. In other cases, this is not as simple to solve. For example, in Steps $6-7$ we need to reintroduce an empty multiset to match with $\Gamma$ in the $L2\times$ rule through a new subgoal (using \hl{SUBGOAL_TAC}).

Even through this simple example, it is clear that performing such object-level proofs with an encoded logic can be tedious. We need to be able to manipulate the goal state to allow the rules to match, which often requires multiple steps using rewritting or appropriate rule instantiations. This effort put into the management of multiset-based context is also encountered in related work~\cite{power1999working}.

 Our framework facilitates this process, alleviating the need for such fine-grained manipulation. More specifically, it provides both forward and backward reasoning with simple procedural tactics. It also performs intelligent multiset matching automatically, performing all the necessary manipulations (such as adding empty multiset as required) in the background. Using our framework, the proof script from Figure~\ref{fig:hlproof} is simplified as shown in Figure~\ref{fig:hlproofimproved}. Notice how each rule application now only requires a single proof step in HOL Light, using our \hl{ruleseq} command described in Section~\ref{sec:implementation}.

\begin{figure}[htbp]
  \begin{center}
  \begin{tabular}{|r|l|c|}\hline
    Step & Tactic & Goal(s) \\\hline
         & \emph{Initial state} & $\mempty\tstile X\times Y \rightarrow Y\times X $ \\
    1 & \hl{ruleseq ($R\rightarrow$)} & $\msing{X\times Y} \tstile Y\times X $ \\
    2 & \hl{ruleseq ($C$)} & $\msing{X\times Y}\munion \msing{X\times Y} \tstile Y\times X $ \\
    3 & \hl{ruleseq ($R\times$)} & $\msing{X\times Y} \tstile Y \qquad \msing{X\times Y} \tstile X $ \\
    4 & \hl{ruleseq ($L2\times$)} & $\msing{Y} \tstile Y \qquad \msing{X\times Y} \tstile X $ \\
    5 & \hl{ruleseq ($ID$)} & $\qquad\qquad \msing{X\times Y} \tstile X $ \\
    6 & \hl{ruleseq ($L1\times$)} & $\qquad \qquad \msing{X} \tstile X$ \\
    7 & \hl{ruleseq ($ID$)} &  \\\hline
  \end{tabular}
  \end{center}
\hspace*{1cm} \hl{let TIMES\_COMM = prove\_seq ( `$\mempty \tstile\ X \times Y \rightarrow Y \times X$`, }\\
\hspace*{1.5cm} \hl{ETHENL (ETHEN (ruleseq $R\rightarrow$) (ETHEN (ruleseq $C$) (ruleseq $R\times$)))} \\
\hspace*{2cm} \hl{[ ETHEN (ruleseq $L2\times$) (ruleseq $ID$) ;}\\
\hspace*{2cm} \hl{$\ $ ETHEN (ruleseq $L1\times$) (ruleseq $ID$) ] );;}
  \caption{Interactive (top) and packaged (bottom) steps performing proof~\eqref{eq:timescomm} using our framework.}
  \label{fig:hlproofimproved}
\end{figure}

\section{Example: The Curry-Howard isomorphism}%
\label{sec:example:ch}


Howard's seminal paper describes the use of first order logic as a system for simply typed $\lambda$-calculus and studies the correspondence of cut elimination to function evaluation in what came to be known as the Curry-Howard correspondence~\cite{Howard80}. Howard uses natural deduction for his formulation, noting how it is ``inappropriate'' for Gentzen's sequent calculus. Girard delves deeper into this distinction between the two proof systems and provides a translation from natural deduction to intuitionistic sequent calculus (i.e.\ sequents with a single conclusion)~\cite{Girard:1989:PT:64805}. 

To simplify the discussion, we focus on a subset of this type theory involving only function and product types as primitives, to mirror the simple logic presented in the previous section, but also in the spirit of Wadler's discussion of the same topic~\cite{Wadler:2015:PT:2847579.2699407}. The corresponding $\lambda$-calculus is defined with the following syntax:

\[
  Lambda\ ::=\ Var\ A\ |\ Lambda\ Lambda \ |\ (Lambda, Lambda)\ |\ \lambda A.\ Lambda 
\]

This includes variables ($Var$), function application ($f\ x$), products $(x,y)$, and functions ($\lambda x.\ y$). Note that this type definition is polymorphic with respect to the type of variables $A$. This allows us to describe the (simple) types of this calculus using any datatype (though typically one would use strings). 

We also define two projection functions for products $fst$ and $snd$ so that $fst\ (x,y) = x$ and $snd\ (x,y)=y$. Normally, projections are added as a primitive constructor of the inductive type. Cut elimination can then be used to prove what we give by definition here. Since we will not be performing cut elimination proofs, we employ these definitions as a more pragmatic approach.

Using the above syntax and mirroring the rules shown in Figure~\ref{fig:simplelogic}, the rules of this type theory are shown in Figure~\ref{fig:curryhoward}.

\begin{figure}[htbp]
  \centering
  \begin{tabular}{c}
    Identity \& Cut \\\hline
    \begin{minipage}{0.8\linewidth}
      \vspace*{3mm}
      \begin{equation}
        \nonumber
        \infer[Ax]{\seq{x}{A} \tstile \seq{x}{A}}{}
        \qquad\qquad
        \infer[Cut]{\Gamma \sep \Delta \tstile \seq{x}{C}}{
        \Gamma \tstile \seq{z}{A}
        &
        \Delta \sep \seq{z}{A} \tstile \seq{x}{C}
      }
      \end{equation}      
    \end{minipage}\vspace*{5mm}\\
    Structural Rules \\\hline
    \begin{minipage}{0.8\linewidth}
      \begin{equation}
        \nonumber
        \infer[Exchange]{\Gamma \sep \seq{y}{B} \sep \seq{x}{A} \sep \Delta \tstile \seq{z}{C}}{
        \Gamma \sep \seq{x}{A} \sep \seq{y}{B} \sep \Delta \tstile \seq{z}{C}
      }
        \qquad\qquad
        \infer[W]{\Gamma \sep \seq{x}{A} \tstile \seq{z}{C}}{
        \Gamma \tstile \seq{z}{C}
      }
        \qquad\qquad
        \infer[C]{\Gamma \sep \seq{x}{A} \tstile \seq{z}{C}}{
        \Gamma \sep \seq{x}{A} \sep \seq{x}{A} \tstile \seq{z}{C}
      }
      \end{equation}      
    \end{minipage}\vspace*{5mm}\\
    Logical Rules \\\hline
    \begin{minipage}{0.8\linewidth}
      \begin{equation}
        \nonumber
        \infer[L1\times]{\Gamma \sep \seq{x}{A \times B} \tstile \seq{z}{C}}{
        \Gamma \sep \seq{fst\ x}{A} \tstile \seq{z}{C}
      }
        \qquad\qquad
        \infer[L2\times]{\Gamma \sep \seq{x}{A \times B} \tstile \seq{z}{C}}{
        \Gamma \sep \seq{snd\ x}{B} \tstile \seq{z}{C}
      }
        \qquad\qquad
        \infer[R\times]{\Gamma \sep \Delta \tstile \seq{(x,y)}{A\times B}}{
        \Gamma \tstile \seq{x}{A}
        &
        \Delta \tstile \seq{y}{B}
      }
    \end{equation}\vspace*{-5mm}
    \begin{equation}
        \nonumber
        \infer[L\rightarrow]{\Gamma \sep \Delta \sep \seq{f}{A\rightarrow B} \tstile \seq{z}{C}}{
        \Gamma \sep \seq{f y}{B} \tstile \seq{z}{C}
        &
        \Delta \tstile \seq{y}{A}
      }
        \qquad\qquad
        \infer[R\rightarrow]{\Gamma \tstile \seq{\lambda x. y}{A\rightarrow B}}{
        \Gamma \sep \seq{Var\ x}{A} \tstile \seq{y}{B}
      }
      \end{equation}      
    \end{minipage}

		\end{tabular}
			\caption{The inference rules of the Curry-Howard correspondence for our given subset of propositional logic.}
			\label{fig:curryhoward}
\end{figure}

It is worth mentioning the explicit distinction between any lambda term and specifically variables. The $R\rightarrow$ rule enforces that only variables can be entered in a $\lambda$ expression, a property which is usually implicit in such formulations.

In the next sections, we describe how this encoding can be built on top of the previous one in HOL Light, and explain what kind of object-level proofs we would like to perform.

\subsection{Encoding}%
\label{sec:ch:embedding}

The encoding of such a type theory can be very similar to the one we described in Section~\ref{sec:embedding} for the same logic without computational components. The main difference is that in this case, our logical terms are augmented with lambda terms whose types they are describing. For this reason, we define the operator $:$ (represented as $::$ in HOL Light to avoid a conflict with its own type annotation) as an infix term constructor. 

The inference rules are defined inductively in the same way, simply using annotated terms instead of propositions. As an example, the $R\times$ rule is defined as follows:

\begin{equation}
  \label{eq:ch:XR}
  \Gamma \tstile \seq{x}{A} \mimp \Delta \tstile \seq{y}{B} \mimp (\Gamma \munion \Delta) \tstile \seq{(x,y)}{A\times B}
\end{equation}

It is worth noting that in other types of computational correspondence, such as those involving linear logic (see Section~\ref{sec:further}), calculus terms can be attached to the entire sequent (as opposed to formulas within it). We accomplish this by adding an additional argument to the consequence function (i.e.\ to $\tstile$).

\subsection{Object-level proofs}%
\label{sec:ch:proofs}

As mentioned in Section~\ref{sec:introduction}, there are 2 types of proofs that can be performed at the object-level with such a logic: \emph{type checking} and \emph{construction proofs}.

The example proof of commutativity of $\times$ from Section~\ref{sec:proofs} can be used to type check the function $\lambda x. (snd(Var\ x),fst(Var\ x))$ as follows:
\begin{equation}
  \label{eq:ch:timescomm}
  \infer[R\rightarrow]{\tstile \seq{\lambda x. (snd(Var\ x),fst(Var\ x))}{X\times Y \rightarrow Y\times X}}{
    \infer[C]{\seq{Var\ x}{X\times Y} \tstile \seq{(snd(Var\ x),fst(Var\ x))}{Y\times X}}{
      \infer[R\times]{\seq{Var\ x}{X\times Y}\sep \seq{Var\ x}{X\times Y} \tstile \seq{(snd(Var\ x),fst(Var\ x))}{Y\times X}}{
        \infer[L1\times]{\seq{Var\ x}{X\times Y}\tstile \seq{fst(Var\ x)}{X}}{
          \infer[Ax]{\seq{fst(Var\ x)}{X}\tstile \seq{fst(Var\ x)}{X}}{}
        }
        &
        \infer[L2\times]{\seq{Var\ x}{X\times Y}\tstile \seq{snd(Var\ x)}{Y}}{
          \infer[Ax]{\seq{snd(Var\ x)}{Y}\tstile \seq{snd(Var\ x)}{Y}}{}
        }
      }
    }
  }
\end{equation}

Notice how, as far as the logical derivation is concerned, proofs~\eqref{eq:timescomm} and \eqref{eq:ch:timescomm}, i.e.\ the proof with and without computational annotations, are identical (although type checking a different function of the same type would require a different proof). Our framework reflects this property, since the proof script from Figure~\ref{fig:hlproofimproved} can be used \emph{exactly as is} to perform proof~\eqref{eq:ch:timescomm}.

The equivalent construction proof aims to construct the term corresponding to the type $X\times Y \rightarrow Y\times X$ from the proof instead of knowing it a priori. We can express such proofs using existential quantification in the meta-logic (HOL) to set the following goal:
\[  \exists f.\ \tstile \seq{f}{X\times Y \rightarrow Y\times X}
\]

Although type checking proofs can be performed in a relatively straightforward way, using appropriate rule applications that match not only the logical but also the computational terms, \emph{construction proofs} can be challenging to do even on paper. Although the logical derivation relies only on the logical terms, the construction of the computational term requires careful tracking of metavariables across the entire proof. For example, the above proof can only be completed by (iteratively) instantiating $f$ to $\lambda x. (snd(Var\ x),fst(Var\ x))$.

In our framework, the same proof script from Figure~\ref{fig:hlproofimproved} can be used for the construction proof. The construction is performed automatically in the background, as described in Section~\ref{sec:construction}, and the constructed term can be extracted once the proof is complete with a simple instantiation of $f$.

\section{Implementation}%
\label{sec:implementation}

In this Section, we describe the key implemented features of our framework. These include the following:
\begin{enumerate}
\item Adapted procedural tactics for forward and backward reasoning inspired from Isabelle (Section~\ref{sec:tactics}).
\item Smarter matching of sequents using multiset matching (Section~\ref{sec:multisets}) and metavariable unification (Section~\ref{sec:construction}).
\item Updated functions for proof state management and tactic application to allow arbitrary extensions of the proof state (Section~\ref{sec:metavariables}).
\item Better management of metavariables, ensuring they are carried forward in the extended proof state for each subgoal (Section~\ref{sec:metavariables}).
\item Functions that facilitate the extraction of constructed components (Section~\ref{sec:construction}).
\end{enumerate}

The full implementation, including the examples included in this paper in full detail, can be found online. The code base is separated into a library of general purpose tools\footnote{\url{https://github.com/PetrosPapapa/hol-light-tools}} (for instance including the extended tactic system described in Section~\ref{sec:metavariables} and the multiset theorems used in Section~\ref{sec:multisets}), and the logic encoding library itself\footnote{\url{https://github.com/PetrosPapapa/hol-light-embed}}. An online tutorial providing a more hands-on guide to encoding the logics described in this paper and the use of our framework is also provided\footnote{\url{https://petrospapapa.github.io/hol-light-embed/CurryHoward.html}}.

\subsection{Procedural tactics}%
\label{sec:tactics}

Traditional use of HOL systems, such as HOL4 and HOL Light, dictates the use of tactics for backwards reasoning and so called \emph{rules} for forward reasoning~\cite{harrison1996hol}. Tactics are applied on a goal and produce a (possibly empty) set of new subgoals, whereas \emph{rules} are functions that combine one or more theorems or assumptions to derive new facts. One of the main problems with this approach, particularly in the context of an encoded logic, is that every inference rule needs to be expressed both as a new tactic that applies the inference backwards (as we would typically want a lot more flexibility than what is offered by \hl{MATCH\_MP\_TAC}) and as a new rule/function that applies the inference forwards. In our example, this would require 9 new tactics and 9 new rules/functions for the primitive inference rules of our simple logic in Figure~\ref{fig:simplelogic} (except the $Exchange$ rule), plus a new tactic and a new function for each derived rule. Our Curry-Howard encoding from Figure~\ref{fig:curryhoward} would require a different implementation of another 9 tactics and 9 rules.

An alternative, more flexible approach can be found in the procedural proof tactics for natural deduction in Isabelle~\cite{paulson1994igt}, namely \hl{rule}, \hl{erule}, \hl{drule} and \hl{frule}. These enable the usage of any arbitrary theorem in a proof, either as a forward reasoning step (manipulating assumptions -- \hl{drule} and \hl{frule}), as a backwards reasoning step (breaking down the goal -- \hl{rule}), or simultaneous forward and backwards reasoning (\hl{erule}). They essentially separate the mechanism of matching a rule to a particular goal state and the rule itself.
There are also the four alternatives \hl{rule\_tac}, \hl{erule\_tac}, \hl{drule\_tac}, and \hl{frule\_tac}, which can be used to partially instantiate a rule before applying it (for example in order to resolve ambiguities when a rule can be matched to a particular proof state in more than one way). Using these tactics, one can manipulate custom inference rules from any encoded logic. 

For this reason, we make use of our \emph{Isabelle Light} framework~\cite{papapanagiotou2010isabelle}, which emulates the aforementioned Isabelle tactics in HOL Light. It also includes a few tactics for managing metavariables, which are key to this work (see Section~\ref{sec:metavariables}). 

The extension of Isabelle Light with the new features described in this work resulted in a new set of procedural tactics tailored to object level reasoning with encoded sequent calculus logics. The new tactics with the additional functionality are marked with a ``\hl{seq}'' tag in their name. For example, \hl{ruleseq} is the extension of \hl{rule}, whereas \hl{rule\_seqtac} is the extension of \hl{rule\_tac} (and similarly for \hl{erule}, \hl{drule}, and \hl{frule}).

\subsection{Multiset matching}%
\label{sec:multisets}

In Section~\ref{sec:embedding}, we discussed the use of multisets to represent sequents in order to avoid the \emph{Exchange} rule, followed by examples of the structural manipulation needed in proofs in Section~\ref{sec:proofs}. When using inference rules in our system we aim to match the multisets describing the involved formulas appropriately, with minimum effort from the user.

Let us take the $L1\times$ rule as an example:
\[
      \infer[L1\times]{\Gamma \sep A \times B \tstile C}{
        \Gamma \sep A \tstile C
      }
\]%
When matching the context of the conclusion, i.e.\ $\Gamma \munion \msing{A\times B}$, we need to find matches for the $\Gamma$ multiset and the $\msing{A\times B}$ term. 

Let us consider a goal with context $\msing{x\times y}$ for some $x, y$. Our goal here is to obtain the instantiation $\{A/x,\ B/y\}$ in order to split the conjunction. It is obvious, however, that the multisets do not match directly, as the goal is not a sum $\munion$. Although $\msing{A\times B}$ matches with $\msing{x\times y}$, there is no component to match with $\Gamma$. Instead, we need to introduce an empty multiset so that the goal's context becomes $\mempty\munion\msing{x\times y}$.

If, instead, the context of the goal was $\msing{x\times y}\munion \msing{z}$ for some $z$, the structure would match the expected sum. However, if we match the multisets directly we would obtain the instantiation $\{\msing{x\times y}/\Gamma\}$ and then our match would fail as $z$ and $A\times B$ do not match.

Finally, if the context of the goal was a larger sum, such as $\msing{x\times y}\munion \msing{z}\munion \msing{w}$ for some $w$, then we would want $\Gamma$ to be matched to the whole of $\msing{z}\munion \msing{w}$.

We therefore implement an algorithm that properly matches multiset parts of sequents by incorporating commutativity and associativity of multiset sum $\munion$. Performing this type of \emph{AC-matching} is a well studied problem~\cite{kapur1992complexity}. Our particular case incolves matching of terms with no shared variables (eliminating the need for occurs checks) and no nested AC functions (we only have a single flat level of multisets), making this problem more tractable than the general case and solvable in P. Our algorithm splits the multisets into their elements, which are either singleton multisets (formulas) or multiset variables (environments). Then it performs the following matches:

\begin{enumerate}
	\item First it tries to match elements of the multiset taken from the inference rule that do not contain free variables, i.e.\ constants and terms pre-instantiated by the user.
	\item Then it tries to match elements that are not variable multisets. In our example, $\msing{A\otimes B}$ is a singleton set that will be matched first. 
	\item Multiset variables (such as $\Gamma$ in our example) are left for last because they can match any part of the target. If the target does not have enough elements for all such multiset variables, they are matched to the empty multiset, whereas if there are more elements left in the goal than the available variables, they are combined into a single multiset sum.
\end{enumerate}

If a match is found for all elements of the rule, the rule is instantiated accordingly. We then apply multiset normalisation via rewriting with the properties of multiset sum, which leads to the rule and the target (goal and assumption) to match exactly, thus allowing the appropriate LCF style justification of the rule application. 

As an extra step, we eliminate any remaining empty multisets from the resulting subgoals. These are viewed as part of the internal mechanics of structural manipulation and outside of the encoded logic. They should not be of interest to the user, instead making the proof goal more confusing.

It is worth noting that our algorithm does not currently support backtracking when matching multiset variables. It merely attempts to find one possible match, under the assumption that if the user needs a different match, they will explicitly instantiate the rule (see Section~\ref{sec:tactics}) to guide the algorithm accordingly. Supporting backtracking is non-trivial as it requires an explicit interaction with the proof state and the tactic application mechanisms, but we are considering it as future work as it can prove particularly useful in automated proof procedures.

\subsection{Metavariables}%
\label{sec:metavariables}

Metavariables in a proof are variables whose instantiations are deferred to later stages in the proof or until the proof is finished. They are particularly useful in a \emph{construction} proof, where the computational translation (i.e.\ the $\lambda$-calculus term in our example) is initially unknown and constructed during the proof (see Section~\ref{sec:ch:proofs}). It is worth noting that, although metavariables exist as a feature in HOL Light, they are rarely used and are seen as ``a bit of a historical accident''\footnote{Personal communication with John Harrison, 2009.}. However, in our case they are essential in order to construct computational translations.

Metavariables are shared by all subgoals of a proof, and thus HOL Light stores them beyond the scope of any single subgoal. In contrast, HOL Light tactics are functions that apply to a single subgoal. As a result, tactics have no access to any information on \emph{already existing} metavariables. This causes problems during the application of our tactics, since they need to (a) be able to freely instantiate metavariables to anything that matches and (b) ensure that metavariables that are newly introduced during construction are fresh.

We address this issue by enhancing HOL Light so that tactics can handle subgoals extended with additional information about the proof state (as a state monad). We call these extended tactics \texttt{etactics} and their definition, contrasted to the definition of a regular HOL Light \texttt{tactic} is shown below:

\begin{verbatim}
        type tactic = goal -> goalstate ;;
        type 'a etactic = 'a -> goal -> (goalstate * 'a) ;;
\end{verbatim}

Note that \texttt{'a} is a type variable that allows the state to be of any type. In our case, this extension allows for our tactics to be given (a) an integer counter that ensures freshness of new variables (similarly to the \texttt{genvar} mechanism of HOL Light) and (b) the list of currently used metavariables, in addition to the target subgoal. 

The functions \texttt{eseq}, \texttt{prove\_seq}, \hl{ETHEN}, \hl{ETHENL}, \hl{EORELSE}, etc.\ shown in our examples are part of this extension, are therefore applicable to \texttt{etactics}, and correspond to the HOL Light functions \texttt{e} (apply a tactic interactively), \texttt{prove} (prove a lemma with a composite tactic), \hl{THEN} (sequential application of tactics), \hl{THENL} (apply a tactic and then apply a list of tactics, each to one of the produced subgoals), \hl{ORELSE} (try to apply a tactic and if it fails apply a different tactic), etc. Note that, the original HOL Light tactics can be used within this extension with the \hl{ETAC} keyword:

\begin{verbatim}
        let (ETAC: tactic -> 'a etactic) = fun tac s g -> tac g,s ;;
\end{verbatim}

Another challenge in the effective use of metavariables lies in the justification of every rule application. The LCF approach adopted in HOL Light requires that every tactic application can formally reproduce the original goal from its generated subgoals. Metavariables make this more challenging as the original goal and subgoals may mutate at any point based on the instantiation of their metavariables. The justification mechanisms for \hl{ruleseq} and the other tactics successfully deal with this issue. Moreover, our efforts on this front uncovered a 20 year old bug with metavariable instantiations in HOL Light\footnote{\url{https://github.com/jrh13/hol-light/pull/52}}.

\subsection{Construction}%
\label{sec:construction}

\emph{Computational construction} in HOL Light proofs is enabled through the use of metavariables. Unknown computational components (i.e.\ $\lambda$-calculus annotations in our example) are treated as metavariables that are gradually instantiated through the application of the inference rules. Note that when using an inference rule, any variables that are not matched to any (sub)terms of the assumptions or the goal are added as metavariables in the new subgoals. At the end of the proof, applying the resulting instantiations to the initial goal's metavariables gives us the computational terms resulting from that proof. 

To demonstrate how such construction proofs work in our framework, let us return to the proof of commutativity of $\times$ discussed in Section~\ref{sec:ch:proofs}:

\begin{equation}
  \label{eq:constr:timescomm}
  \exists f.\ \tstile \seq{f}{X\times Y \rightarrow Y\times X}
\end{equation}

First, we use the HOL Light \hl{META\_EXISTS\_TAC} tactic to eliminate the existential quantifier and add the variable $f$ to the list of metavariables, which allows it to be instantiated gradually in the attempt to find the correct witness. We then perform a backwards proof, by applying the associated inference rules. The first step, involves the $R\rightarrow$ rule:
\[
        \infer[R\rightarrow]{\Gamma \tstile \seq{\lambda x. y}{A\rightarrow B}}{
        \Gamma \sep \seq{Var\ x}{A} \tstile \seq{y}{B}
      }
\]

The backwards application of the rule (using \texttt{ruleseq}) tries to match the conclusion of the rule $\Gamma \tstile \seq{\lambda x. y}{A\rightarrow B}$ to the current goal $\tstile \seq{f}{X\times Y \rightarrow Y\times X}$. Knowing that $f$ is a metavariable, the framework uses \emph{unification} instead of matching. This yields the following instantiation:

\begin{equation}
  \label{eq:constr:i1}
  \{\lambda x. y/f,\ X\times Y/A,\ Y\times X/B,\ \mempty/\Gamma\}  
\end{equation}

\noindent Note that $x$ and $y$ are variables in the rule that are not matched to any subterms in the goal, so they are added as metavariables in the new goal $\seq{Var\ x}{X\times Y} \tstile \seq{y}{Y\times X}$.

The contraction ($C$) step simply yields the goal $\seq{Var\ x}{X\times Y}\sep \seq{Var\ x}{X\times Y} \tstile \seq{y}{Y\times X}$. We then want to apply the $R\times$ rule (with all variables primed for freshness):
\[
 \infer[R\times]{\Gamma' \sep \Delta' \tstile \seq{(x',y')}{A'\times B'}}{
        \Gamma' \tstile \seq{x'}{A'}
        &
        \Delta' \tstile \seq{y'}{B'}
      }
\]

\noindent Following the same unification mechanism as before, we obtain the new instantiation:
\begin{equation}
  \label{eq:constr:i2}
\{(x',y')/y,\ Y/A',\ X/B',\ \msing{\seq{Var\ x}{X\times Y}}/\Gamma',\ \msing{\seq{Var\ x}{X\times Y}}/\Delta'\}
\end{equation}

At this point, we can observe the gradual instantation of $f$, by composing the 2 instantations~\eqref{eq:constr:i1} and~\eqref{eq:constr:i2}, yielding $\{\lambda x. (x', y')/f\}$.

The rest of the proof yields further instantations of the metavariables $x'$ and $y'$, eventually resulting in the final instantation for $f$, namely $\lambda x. (snd(Var\ x),fst(Var\ x))$. It is worth noting that $x$ is still a metavariable in this final result, having never been instantiated to anything else in the proof.

This process demonstrates a fruitful use of metavariable unification and instantiation to allow construction proofs. This is only achieveable thanks to the multiset matching and metavariable management functionality described in the previous sections.

One of the key benefits of this approach is that the user never deals with any of the computational terms explicitly. In fact, one can install a custom term printer in HOL Light to hide the $\lambda$ terms completely from the proof, whilst the proof script remains unaffected. This is in-line with the natural expectation that the logical proof should not be affected by the computational annotations.

In addition to this, we have introduced a \hl{constr\_prove} command to further facilitate construction proofs. As previously mentioned, construction proofs produce existenatially quantified lemmas, such as~\eqref{eq:constr:timescomm}. These are not very practical, as the constructed term is not visible in the lemma, but can only extracted through the metavariables in the finished proof state. The \hl{constr\_prove} command proves the existentially quantified goal, then strips the quantifiers and replaces the variables with the constructed terms. This results in a proven theorem with all the constructed components instantiated.

For instance, consider the following command, which proves lemma~\eqref{eq:constr:timescomm} using a packaged version of the proof script from Figure~\ref{fig:hlproofimproved}:

\vspace{2mm}
\hl{constr\_prove ( `$\exists f.\ \mempty\tstile\ \seq{f}{X\times Y \rightarrow Y\times X}$`, }\\
\hspace*{1.5cm} \hl{ETHENL (ETHEN (ruleseq $R\rightarrow$) (ETHEN (ruleseq $C$) (ruleseq $R\times$)))} \\
\hspace*{2cm} \hl{[ ETHEN (ruleseq $L2\times$) (ruleseq $ID$) ;}\\
\hspace*{2cm} \hl{$\ $ ETHEN (ruleseq $L1\times$) (ruleseq $ID$) ] );;}
\vspace{2mm}

\noindent This command will yield the theorem $\mempty \tstile \seq{\lambda x. (snd(Var\ x),fst(Var\ x))}{X\times Y \rightarrow Y\times X}$ directly.

In addition to making the user workflow for producing constructed lemmas easier, this facilitates the development of libraries of lemmas for different correspondences of the same logic, because the computational terms associated with each lemma do not need to be explicit. For example, if one chose to use a different correspondence of our simple logic, other than the Curry-Howard correspondence to $\lambda$-calculus, they would need to change the original encoding, by adapting the inductive definition of $\tstile$. However, the proof involving \hl{constr\_prove} shown above would still be valid and it would automatically generate the correct theorem based on the new correspondence.

\subsection{Usage and Integration with HOL Light}%
\label{sec:hol-light}

The implemented generic tactics integrate well within the (extended) tactic system of HOL Light. More specifically, they can be used at different levels, such as the following:

\begin{enumerate}
\item \textbf{Interactively}: They allow the application of inference rules of the encoded logic in a step-by-step interactive proof setting, using the \texttt{eseq} command.

\item \textbf{In packaged proofs}: HOL Light proofs are traditionally packaged in a \hl{prove} statement by combining the tactics that achieve the proof using the so-called \emph{tacticals} such as \hl{THEN} and \hl{EVERY}. Our introduced tactics can also be combined in the same way, with the extended tacticals \hl{ETHEN}, \hl{EEVERY}, etc. For example, the packaged verification proof of the commutativity of $\times$ is shown at the bottom half of Figure~\ref{fig:hlproofimproved}.

\item \textbf{Programmatically}: Our procedural tactics facilitate the construction of advanced procedures for proofs in the encoded logics. For instance, they can be directly used within proof search algorithms, helping to reduce the overhead of manipulating the structure of the sequents. Our implementation includes an example of such an automated proof procedure of a type of Linear Logic proofs (see Section~\ref{sec:further}).

\item \textbf{Visually}: In related research, our tactics have proven useful in the context of diagrammatic reasoning~\cite{Papapanagiotou2012}. In this, the user performs gestures in a purely graphical interface. Each gesture triggers a reasoning task in our encoded Linear Logic, and this is accomplished via the framework described in the current work.
\end{enumerate}

\section{Further Examples: Linear Logic}%
\label{sec:further}

So far, our discussion revolved around a simple subset of propositional logic and its well studied correspondence to $\lambda$-calculus. The usefulness of our framework becomes more apparent when one considers that it can work in the exact same way with the encoding of any logic or correspondence, without the need for any further configuration or metadata of any kind. 

A particular logic, which has multiple evolving correspondences, is linear logic~\cite{girard1995linear}, a substructural logic with no weakening or contraction rules. In the 90s, Abramsky, Bellin and Scott developed the \emph{proofs-as-processes} paradigm~\cite{abramsky1994proofs,bellin1994}, introducing a correspondence between Classical Linear Logic and the $\pi$-calculus~\cite{milner94polyadic} and forming a type system for deadlock-free concurrent processes. We have used this paradigm in conjunction with our reasoning framework extensively for the specification and composition of workflow processes using Classical Linear Logic~\cite{LOPSTR2019}, with real-world applications in the modelling of clinical pathways in the healthcare domain~\cite{jbit13,Papapanagiotou2015}.

As the use of concurrent systems has scaled up dramatically in the past decade, reasoning about their properties, including attempts to ensure deadlock freedom and session fidelity, has been a major research track in concurrency theory for the past years and is an ongoing effort. This effort has brought forth the emergence of new correspondences of linear logic to \emph{session types}~\cite{honda1993}, which are used to provide richer semantics for the types of communicated values and session-based protocols in mobile processes~\cite{gay2003session}. 

Caires et al.\ use Intuionistic Linear Logic terms to describe session types and attach them via a proofs-as-processes style to $\pi$-calculus channels~\cite{caires2010}. Subsequent published papers describe further developments of this theory, including a version for asynchronous communication~\cite{deyoung2012}, a comparison to a Classical Linear Logic based version~\cite{Caires2012}, and the use of dependent session types to describe properties about the information being communicated~\cite{Toninho2011,pfenning2011}. In parallel to this track, Wadler developed a \emph{propositions-as-sessions} theory~\cite{wadler2014}. In it, he chooses to loosen the connection to the original $\pi$-calculus by introducing a new process calculus named \emph{CP}. Reductions in CP are \emph{defined} based on cut-elimination steps in Classical Linear Logic, instead of being defined separately and then having their correspondence to cut-elimination proven. Further developments in the correspondence between linear logic and session types are being produced to this day~\cite{vanHeuvel2020}.

Our framework has the capacity to work hand-in-hand with the meta-theoretic efforts in these strands of work, by providing a formal setting to produce object-level proofs. This can help generate actual process instances using linear logic proofs and examine their properties. For instance, this can shine light into how new session type languages, such as Wadler's CP, can behave in practice.

To demonstrate this, our implementation\footnote{\url{https://github.com/PetrosPapapa/hol-light-embed/blob/master/Examples/}} includes 2 example encodings of linear logic. The first, is an encoding of Intuitionistic Linear Logic with no correspondence. The second is an encoding of the \emph{propositions-as-sessions} paradigm, i.e.\ a correspondence of Classical Linear Logic to CP. The latter includes an automated tactic that can prove a type of linear logic sequents within this encoding.

It is worth noting that both these examples and the Curry-Howard encoding described in this paper can all be loaded in HOL Light at the same time, and managed by the same set of tactics we discussed.

\subsection{Related Work}%
\label{sec:related-work}


There are strong similarities between our system and the core design of Isabelle~\cite{paulson2000isabelle,paulson1994igt}, which provides support for a variety of formal theories, such as Higher Order Logic (\emph{Isabelle/HOL}) and Zermelo-Fraenkel set theory (\emph{Isabelle/ZF}). Essentially, it exposes an intuitionistic meta-logic (\emph{Isabelle/Pure}) that enables the encoding of different theories, and the development of sophisticated proof tactics and procedures for both interactive and automated theorem proving within those theories. Our work draws inspiration from this approach and enables similar functionality (for instance our encoding of Intuitionistic Linear Logic\footnote{\url{https://github.com/PetrosPapapa/hol-light-embed/blob/master/Examples/ILL.ml}} strongly resembles the one in Isabelle\footnote{\url{https://isabelle.in.tum.de/dist/library/Sequents/Sequents/ILL.html}}). The key difference lies in the ability to construct an initially unknown computation of a given type using any given set of primitive logic/type rules annotated with their computational translation. Simple synthesis using Constructive Type Theory has been shown to be possible in Isabelle\footnote{\url{https://isabelle.in.tum.de/dist/library/CTT/CTT}}, but has not been investigated to any depth since its formalisation almost 30 years ago by Paulson. Moreover, we use HOL Light's Higher Order Logic as the meta-logic which is more expressive than Isabelle, but narrows the kinds of logics whose use we can automate. 

We employ a similar perspective in terms of multiset sequents to embed logics to that of Dawson and Gor{\'e}'s implementation in Isabelle/HOL~\cite{dawson2010}. While their work is further evidence of the need and usefulness of frameworks for arbitrary encoded logics, it focuses on reasoning about meta-theoretic properties. Instead, our framework is designed to tackle object-level reasoning and computational construction at that level.

Other systems that are known to rely on the propositions-as-types paradigm such as Coq~\cite{bertot2004interactive} and Agda~\cite{agda} have efficient, specialised procedures to deal with computational construction, but these are exclusive to their particular underlying type systems. They are not generic and so cannot be used to reason with different logics or correspondences.
 
Another similar approach is that of $\lambda$Prolog, a logic programming-based system aimed at hosting encodings of different logics and calculi~\cite{lprolog}. For example, Forum is a linear logic based meta-logic built on top of $\lambda$Prolog~\cite{MILLER1996201}. Although $\lambda$Prolog allows more expressive encodings and provides proof search by default (based on its logic programming backend), it cannot support construction proofs because of its lack of support for metavariables (as argued by Guidi et al.\ who suggest potential extensions to achieve this~\cite{guidi:hal-01410567}). Moreover, $\lambda$Prolog does not provide the guarantees of correctness of modern theorem provers, e.g.\ via the LCF approach~\cite{gordon1980}, as is the case here with HOL Light.


In principle, other logical frameworks, such as Isabelle and Coq, would allow a similar implementation, and we believe some of the ideas and techniques presented here could transfer over. We chose HOL Light because it is a powerful system that allows interaction at a low level, resulting in
much flexibility and programmability. For instance, it allows direct access to the proof structure and involved terms, which facilitates the development of sophisticated custom tactics or even an entire extension of the proof system as described in Section~\ref{sec:metavariables}.

\section{Conclusion}%
\label{sec:conclusion}

In summary, we presented a generic framework for object-level reasoning with encoded logics within HOL Light. We assume the inference rules of an encoded logic are encoded in a sequent calculus style, through the definition of logical inference as a function of logical terms.We focus on sequents that consist of multisets of terms, allowing their arbitrary ordering, though this is not necessarily a requirement.

The framework then exposes tactics, originally inspired by Isabelle's procedural natural deduction tactics, that allow intuitive forward and backward application of these rules. They automatically take care of structural reasoning and appropriate construction of the computational component. The latter provides the means for extracting correct-by-construction terms via any correspondence in the style of Curry-Howard. The tactics rely on a complex implementation involving multiset matching, unification, and metavariables. Although they work within an extended tactic system, they can integrate with the HOL Light proof environment, and can be used interactively, in packaged proofs, programmatically (for example to construct automated proof procedures), or even visually. 

We demonstrated the functionality of the tactics through a simple propositional logic and its $\lambda$-calculus correspondence. Encodings of linear logic and its correspondence to the $\pi$-calculus or to session types, two of which are included as examples in the framework, showcase the usefulness of this work towards the study and development of deadlock-free software with session types.

Future goals include optimisations in terms of efficiency and metavariable management. We will also study the encoding of quantified object logics, as part of a more in-depth consideration of the full capabilities and limitations of the framework. Moreover, we are exploring ways to support natural deduction style rules in addition to the currently supported sequent calculus style. 

We believe our implementation provides the basis for facilitated encodings of logics within HOL Light, without the need to reimplement an entire proof engine from scratch. This creates the potential to easily and effectively test the behaviour of different logics and their computational translations. Based on this, our framework can prove to be powerful tool for research in type theory and programming languages, in addition to its already successful application in process specification and composition and the healthcare domain.

\section*{Acknowledgement}

We would like to thank the attendees of Logical Frameworks and Meta Languages:
Theory and Practice (LFMTP) 2020 and the anonymous reviewers for their constructive feedback.

\bibliographystyle{eptcs}
\bibliography{paper}
\end{document}